\documentclass[showpacs,preprintnumbers,onecolumn,12pt]{revtex4}
\usepackage{amssymb}
\usepackage{amsfonts}
\usepackage{amsmath}
\usepackage{graphicx}
\usepackage{dcolumn}
\usepackage{bm}

\input{tcilatex}

\begin{document}

\title{Chaotic dynamics of charged particles in the field of a finite 
\textit{non-uniform }wave packet}
\author{Ricardo Chac\'{o}n}
\affiliation{Departamento de Electr\'{o}nica e Ingenier\'{\i}a Electromec\'{a}nica,
Escuela de Ingenier\'{\i}as Industriales, Universidad de Extremadura,
Apartado Postal 382, E-06071 Badajoz, Spain}
\date{\today}

\begin{abstract}
A generalization of the Chirikov-Taylor model is introduced to study the
dynamics of a charged particle in the field of an electrostatic wave packet
with an \textit{arbitrary but finite} number of harmonics. The dependence of
both the edge of chaos (dissipative regime) and the deterministic diffusion
on the wave packet width is predicted theoretically and confirmed
numerically, including the case of relativistic particles. Diverse
properties of the standard maps are shown to be non-universal in the
framework of the wave-particle interaction, because these maps correspond to
an \textit{infinite} number of waves.
\end{abstract}

\maketitle





During the past quarter of century the Hamiltonian [1] and dissipative [2]
versions of the so-called standard map (SM) have been widely studied as
basic models of the dynamics of charged particles in the field of an
electrostatic wave packet (see, e.g., refs. [3-4]): 
\begin{equation}
\overset{..}{x}+\gamma \overset{.}{x}=-\frac{e}{m_{e}}\sum_{n=-N}^{N}E_{n}%
\sin \left( k_{n}x-\omega _{n}t\right) ,  \tag{1}
\end{equation}%
where $E_{n},k_{n}$, and $\omega _{n}$ are the amplitudes, wave numbers, and
frequencies, respectively, of the $\left( 2N+1\right) $ plane waves, and $%
\gamma ,e$ and $m_{e}$ are the damping coefficient, the charge and the mass
of the particle, respectively. Specifically, the SM describes the particular
case of an infinite set of waves having the \textit{same} amplitudes, same
wave numbers, and integer frequencies. While its diffusion properties $%
\left( \gamma \equiv 0\right) $ have been shown to be non-universal in the
sense that it does not generalize to a wave spectrum with uncorrelated
phases [5], its assumption of an infinite and uniform amplitude distribution
also seems quite restrictive since it does not permit one to study the
sensitivity of the dynamics to changes in the wave packet width. Physically,
this sensitivity yields changes in the properties and structures of the
phase space, as can be appreciated when comparing, e.g., the cases of two
waves [6-9] and infinite waves [4]. In this Letter a generalized model of
the wave packet structure is introduced to take into account such a \textit{%
finite-size} effect on the particle dynamics. Specifically, it is assumed
that $k_{n}=k_{0},\omega _{n}=\omega _{0}+n\Delta \omega ,E_{n}=E_{n}\left(
m\right) \equiv E_{0}\func{sech}\left[ n\pi K(1-m)/K(m)\right] $, where $K(m)
$ is the complete elliptic integral of the first kind, i.e., a $\func{sech}$
distribution is assumed for the amplitudes such that the effective width is
controlled by a \textit{single} parameter: the elliptic parameter $m$. This
specific form of $E_{n}\left( m\right) $ is motivated by the following
properties: $(i)$ $E_{n}(m=0)=E_{0}\delta _{n0}$, with $\delta _{n0}$ being
the Kronecker delta, i.e., one recovers the (non-chaotic) limiting case of a
single plane wave. $(ii)$ $E_{n}(m=1)=E_{0},\forall n$, i.e., one recovers
the limiting case described by the SM. $(iii)$ For any $m\in \lbrack 0,1)$,
one may define an \textit{effective} number of harmonics forming the wave
packet as follows. Let us choose quite freely a real number $\zeta \in
\left( 0,1\right) $ such that $N_{eff}$ is the largest integer satisfying $%
E_{N_{eff}}/E_{0}\geqslant \zeta $, then $E_{n}/E_{0}<\zeta ,\forall
n>N_{eff}$, i.e., $N_{eff}=N_{eff}(m)\equiv \left[ K(m)\cosh ^{-1}\left(
1/\zeta \right) /(\pi K(1-m))\right] +1$ where the brackets stand for the
integer part. Thus, for this choice of the wave packet structure, one has $%
\sum_{n=-N}^{N}E_{n}\sin \left( k_{n}x-\omega _{n}t\right) =\sin \left(
k_{0}x-\omega _{0}t\right) \sum_{n=-N_{eff}}^{N_{eff}}E_{n}(m)\cos \left(
n\Delta \omega t\right) $ and, after extending the summation from $-\infty $
to $\infty $, eq. (1) transforms into the form%
\begin{equation}
\overset{..}{x}+\gamma \overset{.}{x}=-\frac{eE_{0}}{m_{e}}\sin \left(
k_{0}x-\omega _{0}t\right) D(t;T,m),  \tag{2}
\end{equation}%
where $T\equiv 2\pi /\Delta \omega $ is the characteristic period of the
field and $D(t;T,m)\equiv 2K(m)\limfunc{dn}\left[ 2K(m)t/T;m\right] /\pi $
[10], with $\limfunc{dn}$ being the Jacobian elliptic function of parameter $%
m$. Note that this approximation does not imply a loss of generality \textit{%
at variance with} the case of the SM. In a reference frame moving with the
main wave, eq. (2) transforms into the equation 
\begin{equation}
\frac{d^{2}\xi }{d\tau ^{2}}+\sin \xi =-\delta -\eta \frac{d\xi }{d\tau }-%
\left[ D\left( \tau ;\alpha ,m\right) -1\right] \sin \xi ,  \tag{3}
\end{equation}%
where $\xi \equiv k_{0}x-\omega _{0}t$, $\Omega _{0}\equiv \left(
ek_{0}E_{0}/m_{e}\right) ^{1/2}$, $\tau \equiv \Omega _{0}t$, $\delta \equiv
\gamma \omega _{0}\Omega _{0}^{-2}$, $\eta \equiv \gamma \Omega _{0}^{-1}$,
and $\alpha \equiv \Omega _{0}T$ are all dimensionless variables and
parameters. It is worth noting that $\alpha \equiv \sqrt{k}$, where $k$ is
the stochasticity parameter of the SM [9]. The parameters $k_{0},\omega
_{0},E_{0},$ and $\Omega _{0}$ are held constant throughout. Physically, eq.
(3) represents a damped pendulum subjected to a periodic string of \textit{%
finite} pulses having an effective width and an amplitude controlled by $m$.
Thus, model (2) permits one to study the structural stability of the system
under changes in the width of the wave packet by \textit{solely} varying the
parameter $m$ (and hence $N_{eff}$) between $0$ and $1$. \textit{In
particular}, the case of a single plane wave $\left( m=0\right) $ is
described by a purely damped pendulum, while the case of an infinity of
plane waves (SM: $m=1$) is described by a delta-kicked rotator. The
dependence of the chaotic dynamics' features on the effective number of
harmonics will be considered here.

\emph{Dissipative Regime:} Consider first the case of weak dissipation $%
\left( 0<\gamma \ll 1\right) $. Since $D\left( \tau ;\alpha ,m\right)
-1=(2/E_{0})\sum_{n=1}^{\infty }E_{n}(m)\cos \left[ 2n\pi \tau /\alpha %
\right] $, eq. (3) may be regarded as a perturbed pendulum $\left( 0<\delta
,\eta \ll 1\right) $ for $m\in \lbrack 0,1)$ and then one can apply
Melnikov's method (MM) [11-13,3] to obtain an analytical estimate of the
edge of chaos in the parameter space. The application of MM to eq. (3) gives
the Melnikov function $M^{\pm }\left( \tau _{0}\right) =-D^{\pm }+\left(
16\pi ^{3}/E_{0}\right) \sum_{n=1}^{\infty }n^{2}E_{n}(m)b_{n}(\alpha )\sin
\left( 2n\pi \tau _{0}/\alpha \right) $, with $D^{\pm }=8\eta \pm 2\pi
\delta $, $b_{n}\left( \alpha \right) \equiv \alpha ^{-2}\func{csch}\left(
n\pi ^{2}/\alpha \right) $, and where the positive (negative) sign refers to
the top (bottom) homoclinic orbit of the underlying conservative pendulum.
It is well known that the simple zeros of the Melnikov function imply
transversal intersections of stable and unstable manifolds (i.e., a
homoclinic bifurcation occurs), giving rise to Smale horseshoes and hence
hyperbolic invariant sets [12]. From $M^{\pm }\left( \tau _{0}\right) $ one
sees that a homoclinic bifurcation (signifying the \textit{possibility} of
chaotic behavior) is guaranteed for trajectories whose initial conditions
are sufficiently close to the separatrix of the underlying conservative
pendulum if $\left\vert 2\eta \pm \pi \delta /2\right\vert <U\left( m,\alpha
\right) \equiv \left( 4\pi ^{3}/E_{0}\right) \sum_{n=1}^{\infty
}n^{2}E_{n}(m)b_{n}(\alpha )$, where $U(m,\alpha )$ is the chaotic threshold
function. It is straightforward to obtain the following properties: (\textit{%
i}) $U(m,\alpha )$ increases, as a function of $m$, as $m$ is increased,
i.e., the possibility of chaos increases as the spectral width is increased;
(\textit{ii}) $U(m,\alpha \rightarrow 0,\infty )=0$, i.e., for any spectral
width, the possibility of chaos diminishes when the small-amplitude
frequency of the non-perturbed equivalent pendulum ($d^{2}\xi /dt^{2}+\Omega
_{0}^{2}\sin \xi =0$) is much higher or much lower than the characteristic
spectral frequency; (\textit{iii}) $U\left( m,\alpha \right) $ presents a
maximum, as a function of the parameter $\alpha $, at $\alpha _{\max }\equiv
\alpha _{\max }\left( m\right) ,\forall m\in (0,1),$ such that $\alpha
_{\max }\left( m\right) $ is a monotonously increasing function and exhibits
the asymptotic behavior $\alpha _{\max }\left( m\rightarrow 1\right) \sim
\ln \left( 1-m\right) ^{-1/2}$ ; and (\textit{iv}) $U(m\rightarrow 1,\alpha
)=4\pi ^{3}\sum_{n=1}^{\infty }n^{2}b_{n}\left( \alpha \right) >U(m,\alpha
),\forall m\in \lbrack 0,1)$, i.e., it is expected that the limiting case $%
m=1$ be maximal with respect to the extension of dissipative chaos in
parameter space. Note that properties (\textit{iii}) and (\textit{iv}) mean
that the dissipative SM is \textit{non-universal }in the framework of the
wave-particle interaction, because this map corresponds to an infinite set
of waves having the same amplitudes, and hence it does not present the
aforementioned maximum. Extensive Lyapunov exponent (LE) calculations of eq.
(3) are coherent with properties (\textit{i})-(\textit{iv}). One typically
finds that the maximal LE, $\lambda ^{+}$, presents a maximum as a function
of $\alpha $ at $\alpha _{\max }^{\ast }\equiv \alpha _{\max }^{\ast }\left(
m\right) $ and that both $\lambda ^{+}(\alpha _{\max }^{\ast })$ and $\alpha
_{\max }^{\ast }$ increase with the spectral width in accordance with the
MM-based predictions (It is worth mentioning that one cannot expect too good
a quantitative agreement between $\alpha _{\max }(m)$ and $\alpha _{\max
}^{\ast }(m)$ because MM is a perturbative method generally related to
transient chaos, while LE provides information concerning solely steady
chaos [12]). Figure 1 shows an illustrative example for three increasing
values of the spectral width.

\emph{Dissipationless Regime}\textit{: }When dissipation is negligible $%
\left( \gamma =0\right) $, system (3) is generated by the Hamiltonian $H(\xi
,p_{\xi },\tau )=p_{\xi }^{2}/2-D\left( \tau ;\alpha ,m\right) \cos \xi $, $%
p_{\xi }\equiv k_{0}\overset{.}{x}/\Omega _{0}$, and MM provides an estimate
of the width of the stochastic layer generated around the unperturbed
separatrix [14]: $d\left( m,\alpha \right) =U(m,\alpha )/2\equiv \left( 2\pi
^{3}/E_{0}\right) \sum_{n=1}^{\infty }n^{2}E_{n}(m)b_{n}(\alpha )$. Thus,
the aforementioned properties of the chaotic threshold function hold for the
width of the stochastic layer and hence one expects an intensification of
the deterministic diffusion as the spectral width is increased. Figures 2(b,
d, f) provide an illustrative sequence for three increasing values of $m$ at 
$\alpha \simeq \alpha _{\max }(m)$, respectively. Moreover, one typically
finds the gradual disappearance of the invariant curves region inside the
separatrix cell (cf. figs. 2(c, b, d, f)). Numerical simulations also
confirm that the stochastic layer exhibits a maximal width as a function of $%
\alpha $ at a value close to $\alpha _{\max }\left( m\right) $. This can be
appreciated in the sequence of figs. 2(a, c, e). Numerical simulations also
indicate that another difference with respect to the SM is that the phase
space of the Hamiltonian $H(\xi ,p_{\xi },\tau )$ is bounded by
Kolmogorov-Arnold-Moser (KAM) tori at \textit{any} values of $\alpha $ and $%
m\in \left( 0,1\right) $, i.e., global stochasticity is not possible for any
arbitrary but \textit{finite} spectral width (notice that, contrary to the
SM, the phase space of the Hamiltonian $H(\xi ,p_{\xi },\tau )$ is not
periodic in $p_{\xi }$). One thus concludes that, in the context of the
wave-particle interaction, such a transition to global stochasticity is a
peculiarity of the SM.

\emph{Relativistic Regime}\textit{: }Similarly to the case of the
relativistic standard map (RSM) [15,16], the relativistic generalization of
model (2) ($\gamma \equiv 0$) is necessary if the acceleration of particles
is sufficiently large. The relativistic equations corresponding to model
(2), $\overset{.}{x}=pc^{2}\left( m_{e0}^{2}c^{4}+p^{2}c^{2}\right) ^{-1/2}$%
, $\overset{.}{p}=-eE_{0}\sin \left( k_{0}x-\omega _{0}t\right) D(t;T,m)$,
where $p$, $m_{e0}$, and $c$ are the momentum and the rest mass of the
particle, and the light velocity, respectively, may be conveniently
transformed into the dimensionless form 
\begin{eqnarray}
\frac{d\xi }{d\tau } &=&\frac{p_{\xi }}{\sqrt{1+\epsilon p_{\xi }^{2}}}%
-\Gamma ,  \notag \\
\frac{dp_{\xi }}{d\tau } &=&-\sin \xi D\left( \tau ;\alpha ,m\right) , 
\TCItag{4}
\end{eqnarray}%
where $\epsilon \equiv \left( \Omega _{0}/k_{0}\right) ^{2}/c^{2}$ is the
relativistic parameter, $p_{\xi }\equiv k_{0}p/(m_{e0}\Omega _{0})$ is the
dimensionless momentum, and $\Gamma \equiv \omega _{0}/\Omega _{0}$. To also
characterize the relativistic dynamics in the coordinate-velocity phase
space, $\left( \xi ,d\xi /d\tau \right) $, eq. (4) is rewritten as a
second-order differential equation: $d^{2}\xi /d\tau ^{2}=-D\left( \tau
;\alpha ,m\right) \sin \xi \left[ 1-\epsilon \left( d\xi /d\tau +\Gamma
\right) ^{2}\right] ^{3/2}$. A relativistic effect is the reduction of the
system symmetry: eq. (4) presents two mirror symmetries with respect to $\xi
=0$ and $p_{\xi }=\Gamma $, respectively, for non-relativistic particles,
while it solely presents the former symmetry for relativistic particles when 
$\Gamma >0$. Another relativistic effect is the modification of the fixed
points existing in the classical (Newtonian) regime: $\left( \xi ,p_{\xi
}\right) =\{\left( 0,\Gamma /\sqrt{1-\epsilon \Gamma ^{2}}\right) ,\left(
\pm \pi ,\Gamma /\sqrt{1-\epsilon \Gamma ^{2}}\right) \}$. Two limiting
cases may be distinguished: $\epsilon \ll \Gamma ^{-2}$ and $\epsilon
\lesssim \Gamma ^{-2}$. Physically, the resonance condition $\epsilon
=\Gamma ^{-2}$ means that the phase velocity of the main wave $\left( \omega
_{0}/k_{0}\right) $ is equal to the velocity of light. In the former case,
when the phase velocity differs significantly from the velocity of light, it
is found numerically that the stochastic motion is restricted to the
neighborhood of the fixed points, such that the corresponding chaotic layers
are bounded by KAM tori at any values of $\alpha $ and $m\in \left(
0,1\right) $ (see figs. 3 and 4). For relativistic particles just beyond the
Newtonian regime $\left( \epsilon \gtrsim 0\right) $, it is found
numerically that the extension of the stochasticity regions in the
coordinate-momentum phase space for relativistic and nonrelativistic
particles, respectively, is approximately the same, while it diminishes
relatively in the coordinate-velocity phase space for relativistic
particles, as can be appreciated in the examples shown in figs. 4 and 3,
respectively. Clearly, this shrinkage is a consequence of the relativistic
constraint $\left\vert d\xi /d\tau \right\vert \leqslant \epsilon
^{-1/2}-\Gamma $ (i.e., $\left\vert \overset{.}{x}\right\vert \leqslant c$).
Note that a proper comparison of the structures appearing in the two phase
spaces would require taking the initial momentum as $p_{\xi }(0)\equiv \left[
(d\xi /d\tau )(0)+\Gamma \right] \left\{ 1-\epsilon \left[ \left( d\xi
/d\tau \right) (0)+\Gamma \right] ^{2}\right\} ^{-1/2}$, where $\left( d\xi
/d\tau \right) (0)$ is the corresponding initial velocity. A preliminary
quantitative estimate of such a relativistic effect in the
coordinate-velocity phase space can be obtained from the equation giving the
first-order relativistic correction: $d^{2}\xi /d\tau ^{2}+\sin \xi =-\left[
D\left( \tau ;\alpha ,m\right) -1\right] \sin \xi +\frac{3}{2}\epsilon
D\left( \tau ;\alpha ,m\right) \left( d\xi /d\tau +\Gamma \right) ^{2}\sin
\xi +O\left( \epsilon ^{2}\right) $. Similarly to eq. (3), this equation may
be considered as a perturbed pendulum for $m\in \lbrack 0,1)$ and, after
applying MM to it, one straightforwardly obtains the following expression
for the width of the stochastic layer: $d_{R}\left( m,\alpha ,\epsilon
\right) =U(m,\alpha ,\epsilon )/2\equiv \left( 2\pi ^{3}/E_{0}\right)
\sum_{n=1}^{\infty }n^{2}E_{n}(m)\left[ b_{n}\left( \alpha \right) -\epsilon
r_{n}(\alpha )\right] $, where $r_{n}\left( \alpha \right) \equiv \left[
2\alpha ^{-2}+n^{2}\pi ^{2}\alpha ^{-4}/2\right] \func{csch}\left( n\pi
^{2}/\alpha \right) $ and $U_{R}(m,\alpha ,\epsilon )$ is the first-order
relativistic threshold function. One obtains that, for fixed $m,\alpha $,
the relativistic width $d_{R}$ decreases as $\epsilon $ is increased, and
that this decrease is ever more noticeable as $m$ is increased. Numerical
results confirm these two predictions, as in the example shown in fig. 3.
Proceeding similarly to obtain a first-order relativistic correction in the
coordinate-momentum phase space by expanding the square root in eq. (4), the
MM yields a \textit{null }correction to the Newtonian width because the
corresponding integrals vanish. Numerical results confirm this prediction,
as in the example shown in fig. 4. As the phase velocity approaches the
velocity of light $\left( \epsilon \lesssim \Gamma ^{-2}\right) $, the
momentum of the fixed points increases continuously, and the KAM torus
limiting the stochastic region from above expands into the higher-momentum
region (see fig. 5(a)). When the resonance condition $\epsilon =\Gamma ^{-2}$
is met exactly, the particles can be accelerated to very high energies,
which is indicated by the vertical traces in fig. 5(b). This mechanism of
particle acceleration [17,18] is also observed in the RSM for the case $%
\omega _{0}=2\pi l/T$, with $l$ being an integer [15,16]. Notice that, far
from the resonance condition, the analysis of the RSM indicated [15,16] a
relativity-induced decrease of the stochasticity in the coordinate-momentum
phase space. However, the above findings clearly show that such a property
does not hold for finite wave packets.

In sum, a generalized model of the dynamics of a charged particle in the
field of an electrostatic wave packet with an arbitrary but finite number of
harmonics has been presented. The dependence of both the chaotic threshold
(dissipative regime) and the deterministic diffusion on the spectral width
was predicted theoretically and confirmed numerically, including the case of
relativistic particles. Several properties of the SM and the RSM were shown
to be non-universal in the framework of the wave-particle interaction,
because these maps correspond to an infinite number of waves.

The author thanks J. I. Cirac, J. A. C. Gallas, and Y. Kivshar for
encouragement and useful comments. The author acknowledges partial financial
support from Spanish MCyT and European Regional Development Fund (FEDER)
program through FIS2004-02475 project.

\subsection{Figure Captions}

\bigskip

Figure 1. Maximal LE vs parameter $\alpha $ for three $m$ values: 1-10$%
^{-4},1-10^{-6}$, and 1-10$^{-8}$, which correspond to $N_{eff}=5,7,$ and 8,
respectively, for $\zeta =0.05$. Vertical arrows indicate the position of
the respective maxima. System parameters: $\delta =\eta =0.1$.

\bigskip

Figure 2. Trajectories in the stroboscopic (for $\tau =n\alpha $ with $%
n=0,...,400$) Poincar\'{e} section $\left( p_{\xi }\text{ vs }\xi /\pi
\right) $ for the system (3) with no dissipation $\left( \delta =\eta
=0\right) $, and (a) $m=0.5,\alpha =1.4$, (b) $m=0.95$, $\alpha =5.95\simeq
\alpha _{\max }(m=0.95)$, (c) $m=0.5,\alpha =5.3\simeq \alpha _{\max
}(m=0.5) $, (d) $m=1-10^{-6}$, $\alpha =11.7\simeq \alpha _{\max
}(m=1-10^{-6})$, (e) $m=0.5,\alpha =7.9$, and (f) $m=1-10^{-14}$, $\alpha
=23.5\simeq \alpha _{\max }(m=1-10^{-14})$. For fixed $\zeta =0.05$, one has 
$N_{eff}=2,3,7,14$ for $m=0.5,0.95,1-10^{-6},1-10^{-14}$, respectively.

\bigskip

Figure 3. Trajectories in the stroboscopic (for $\tau =n\alpha $ with $%
n=0,...,400$) Poincar\'{e} section $\left( d\xi /d\tau \text{ vs }\xi /\pi
\right) $ for the relativistic system (see text), and (a) $m=0.6,\epsilon =0$%
, (b) $m=0.6$, $\epsilon =1/9$, (c) $m=0.9,\epsilon =0$, and (d) $m=0.9$, $%
\epsilon =1/9$. System parameters: $\alpha =5.3,\Gamma =0$.

\bigskip

Figure 4. Trajectories in the stroboscopic (for $\tau =n\alpha $ with $%
n=0,...,400$) Poincar\'{e} section $\left( p_{\xi }\text{ vs }\xi /\pi
\right) $ for the relativistic system (see text), and (a) $m=0.6,\epsilon =0$%
, (b) $m=0.6$, $\epsilon =1/9$, (c) $m=0.9,\epsilon =0$, and (d) $m=0.9$, $%
\epsilon =1/9$. The initial conditions correspond to those used in fig. 4.
System parameters: $\alpha =5.3,\Gamma =0$.

\bigskip

Figure 5. Trajectories in the stroboscopic (for $\tau =n\alpha $ with $%
n=0,...,400$) Poincar\'{e} section $\left( p_{\xi }\text{ vs }\xi /\pi
\right) $ for the relativistic system (see text), and (a) $\epsilon =0.8$,
and (b) $\epsilon =1$. System parameters: $m=0.5,\alpha =5.3,\Gamma =1.$

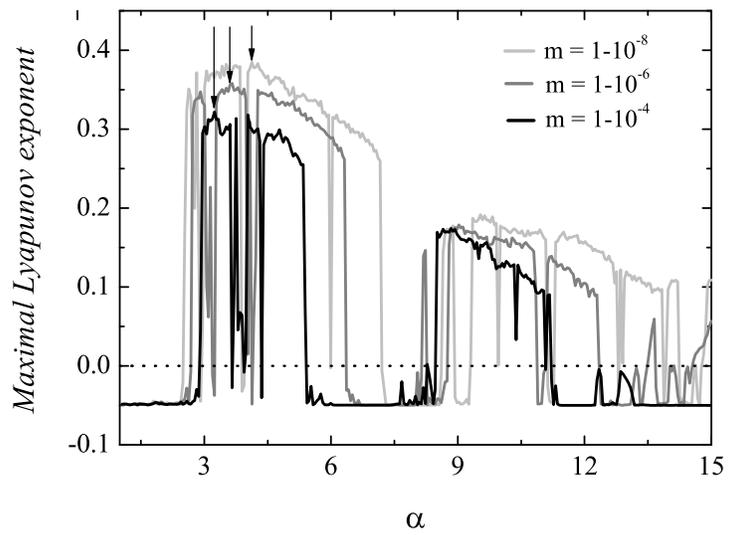
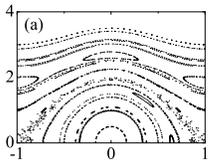
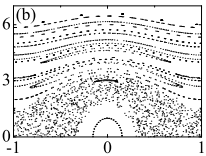
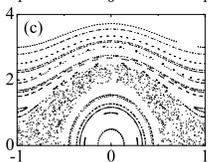
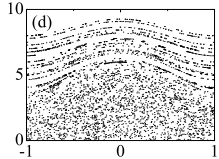
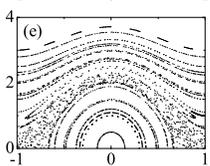
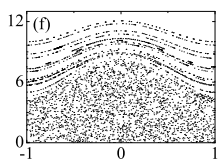
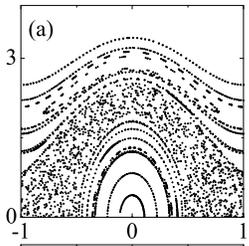
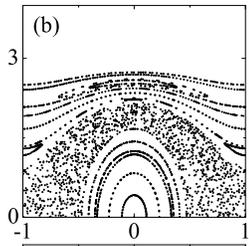
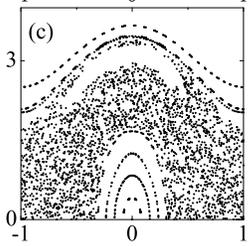
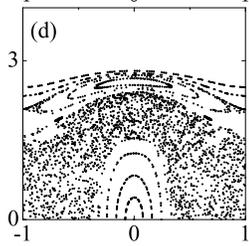
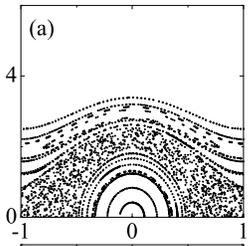
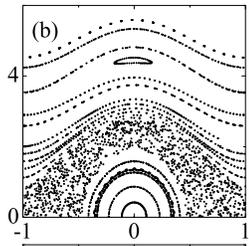
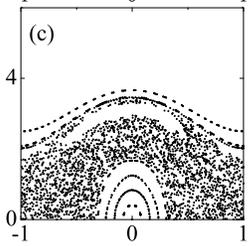
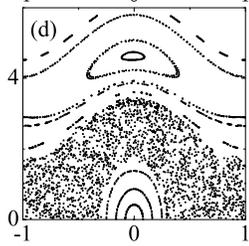
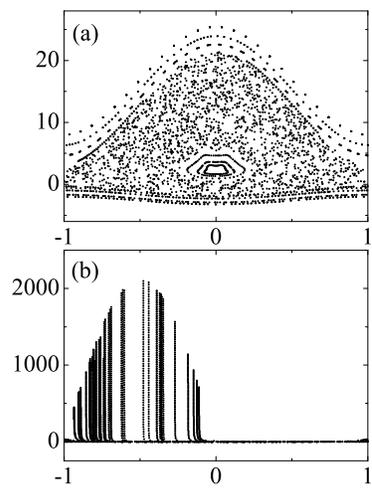
\end{document}